# Mathematical method for calculating batch fragmentations and their impacts on product recall within a FIFO assignment policy

Simon Tamayo[*a], Thibaud Monteiro[b]

[a] MINES ParisTech (CAOR), PSL Research University, 60 boulevard Saint-Michel, 75272 PARIS Cedex 06, France

[b] INSA Lyon (DISP), Université de Lyon, 19 avenue Jean Capelle (Bât Jules Verne), 69621 Villeurbanne Cedex, France

**Abstract**

This study explores the interactions between order sizes, batch sizes and potential product recalls within a FIFO assignment policy. Evidence is provided that the extent of a product recall is related to the fragmentation of the batches of input materials as it amplifies the impact of a crisis. A new management indicator is proposed in order to quantify the expected number of fragments composing a customer order $Fr_{BO}$. A probabilistic analysis reveals that for a given likelihood of crisis, the presence of different batches in a customer order will largely increase its risk. Accordingly, a new equation is proposed for calculating the expected recall size. Taking into account the fragmentation measure allows, for the first time, for the integration of a proactive product recall policy in the batch sizing decision process. A Monte Carlo simulation is performed to validate the effectiveness of this approach.

*Keywords: product recall, batch fragmentation, batch splitting, crisis probability, FIFO, traceability, batch sizing.*

## 1. Introduction

Firms use materials daily to fulfil their customer orders. Typically, they use quantities of a given input to satisfy a requested quantity of an agreed output. This is a general case for different customer-supplier relationships. For instance, in the production process an amount of raw material is transformed to create an amount of finished product in order to manufacture a production order. This is also the case of a distribution operation in which an amount of finished product on-hand is packaged and shipped to satisfy the requested quantity of a delivery order. The input materials are usually organized in batches and it can occur that a customer order is composed of several different batches; if one of such had a crisis (regarding its quality and/or safety standards) then the customer order would be fully recalled.

This work explores the following, for a "first-in first-out" (FIFO) assignment policy: how batch sizes and ordered quantities affect a customer order in terms of risk; how the input batches are split and dispersed downstream though the production and distribution flows; and how such fragmentation influences the likeliness and the impact of a product recall.

It has been proven that product recall announcements have a significant negative impact on the supply chain (Wynn et al., 2011; Hora et al., 2011; Kumar, 2014; Steven et al., 2014), essentially affecting the value of manufacturers and retailers (Zhao et al., 2013; Ni et al. 2014). They need to be accomplished swiftly and flawlessly, or the outcome could be disastrous. For example, every year approximately seven million people are affected by food-borne illnesses in Europe (Kimball and Taneda, 2004). In 2011, the "E. coli crisis" caused 50 deaths in Germany and injured more than 4,000 people (World Health Organization, 2011).

---

[*] Corresponding author.
E-mail addresses: simon.tamayo@mines-paristech.fr (S. Tamayo), thibaud.monteiro@insa-lyon.fr (T. Monteiro).

Significantly, this crisis originated on a single farm, although the propagation of the crisis was such, that before the end of the year the European Commission had allocated €210 million to firms that had lost money due to the disease (Deutsche Welle, 2011).

The rate of product recalls has increased throughout the world during the last decade. The U.S. Food and Drug Administration office (FDA) recorded that the number of recalls related to food and drugs has risen from around 4,700 incidents in 2004 to nearly 8,044 incidents in 2013 (Kumar, 2014). This trend is also present in the industrial setting; for instance, the number of toy-related notifications in Europe rose from 136 in 2003 to 1,803 in 2011 (Memon et al. 2014); in the U.S. the toy-related recall notifications doubled during the same period (Hora et al., 2011).

Flawlessly managing product recalls is an unmistakable objective for any firm today in the manufacturing and retail setting. Traceability enables this ambition as it allows firms to trace (upstream) the batches of input materials and their origin; and track (downstream) customer orders containing such batches (Tamayo et al., 2009). **Yet traceability alone cannot limit the negative impact of recalls, as it merely points to origins and destinations. Proactive action within the operation is therefore crucial to reducing such impacts**.

A beginning of such preventive action is found in the work of Dupuy et al., (2005) as they formalized the notion of raw material dispersion and proposed a mixed integer linear programming model (MILP) for minimizing it in a simplified production layout. Tamayo et al. (2009) addressed more realistic problems (i.e. larger number of variables and constraints) through neural networks and genetic algorithms, obtaining good solutions in reasonable computational time. Following this, Dabbene and Gay (2011) developed a new MILP model for the worst-case and average amount of product recall crisis scenarios. Dhouib et al. addressed similar problems by using other metaheuristics such as record-to-record travel (2010) and artificial bee colonies (2013). Wang (2010) proposed a combined optimization approach of traceability and manufacturing performances, acting on both batch size and batch dispersion, by introducing risk functions.

Despite keen interest in the concept of batch dispersion, the notion of fragmentation has not been explored as a defining factor in the recall crisis scenarios. The concept of batch fragment has been related to traceability and product recalls in terms of identifying difficulties in the food industry (Schwagele, 2005; Ballin, 2010) and the wood industry (ONF, 2004). Olsen and Aschan (2010) related the notion of batch splitting to the loss of information within the materials flow. Such analysis was furthermore adapted to the case of food supply chains (Bosona and Gebresenbet, 2013). To date, the notion of batch fragmentation has however neither been formalized nor related to the production parameters (i.e. scheduling policy, batch sizes, etc.), or associated to the risk in the composed customer orders and the resulting recall outcomes.

In the following sections the notion of recall size and its relation to batch fragmentation are explored to identify some best practices related to batch sizing, in order to effectively reduce the total recalled quantity in case of crisis.

## 2. Problem statement

In the mass production, firms produce mostly in batches. The European Commission defines a batch as **"a quantity of material or product processed in one process or series of processes so that it could be expected to be homogeneous"** (ECDPH, 2004). From this definition it can be stated that all the units or parts of a batch have the same likelihood of meeting a set of quality and/or safety standards, i.e. the entire batch has the same probability of being or not being accepted.

In order to decide which batch (or batches) of materials to use for each agreed order, companies often use a FIFO policy, which is considered to be an inexpensive and beneficial way to manage the materials flow because it minimizes out-dated products and allows items to move quickly (Lee, 2006; Wee and Widyadana, 2013). Depending on the sizes of the batches and orders, the FIFO assignment may split the input materials and – in case of inherent crisis – amplify the risk. In the following section a mathematical analysis is proposed to determine the expected number of fragments (of input batches) composing a customer order, as well as the product recall size for a total ordered quantity.

## 3. Notion of fragmentation

For average customer orders, the batch sizes usually do not correspond to the ordered quantities, batches are therefore frequently fragmented as shown in Figure 1. Three main scenarios may occur: either the batch size is smaller, equal to or greater than the ordered quantity. In Figure 1, the first case (left) presents a batch size smaller than the ordered quantity. The order is composed of batches 1 and 2. Batch 2 is fragmented and it will probably also be present in some other order. The second case (right) shows a batch size greater than the ordered quantity. As a result, the order is entirely composed of batch 1, which is fragmented, and a further order will be composed by two batches: the remainder of batch 1 and the next batch (as stated by the FIFO policy). It is then clear that the fragmentation of batches is a defining and amplifying factor of the presence of different batches, or fragments of batches, within a customer order.

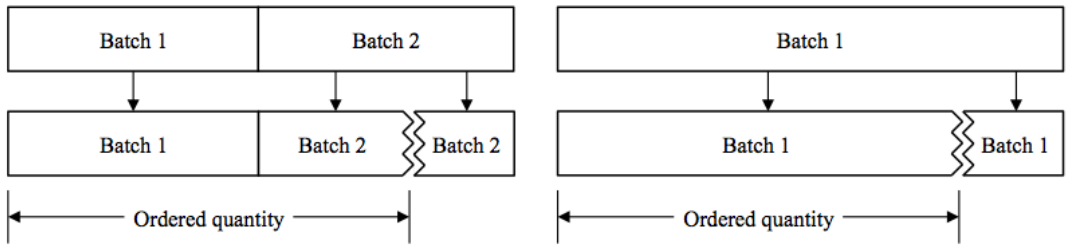

Figure 1. Fragmentation of batches of input materials while fulfilling an ordered quantity.

Should a crisis arise concerning a given batch, all the orders containing it (or a part if it) would be recalled. This is why it is crucial to control the fragmentation of batches during the production and distribution operations.

## 4. Mathematical model

We propose to measure the presence of different batches of a given input on a single customer order and, inversely, the presence of a single batch in different orders, while following a FIFO assignment policy. To do so, we define "$Fr_{BO}$" as the number of fragments of batches "$B$" composing a customer order "$O$". For instance in the production setting, $Fr_{BO}$ defines the number of fragments of different batches of a given raw material that are present in a finished product order. For a crisis probability that can be considered independently for each batch, according to the definition of the European Commission (ECDPH, 2004), the number of batches composing the order will affect its inbuilt risk and furthermore its inherent probability of recall.

### 4.1. Calculating the number of fragments composing a customer order

The value of $Fr_{BO}$ relates to the customer order size and the batch size. The ratio between the two determines the lower and upper boundaries for the values of $Fr_{BO}$ as follows:

$$Fr_{BO} \in \{Fr_{MIN}, Fr_{MAX}\}$$

$$Fr_{BO} \in \left\{ \left\lfloor \frac{O_S}{B_S} \right\rfloor, \left\lfloor \frac{O_S}{B_S} \right\rfloor + 1 \right\}$$

With:

$B_S$ = Input material batch size

$O_S$ = Customer order size

Example: for customer orders of 10 units, if the size of production batches is 4 units, there will be between 3 and 4 batches contained in the order, as shown in Figure 2.

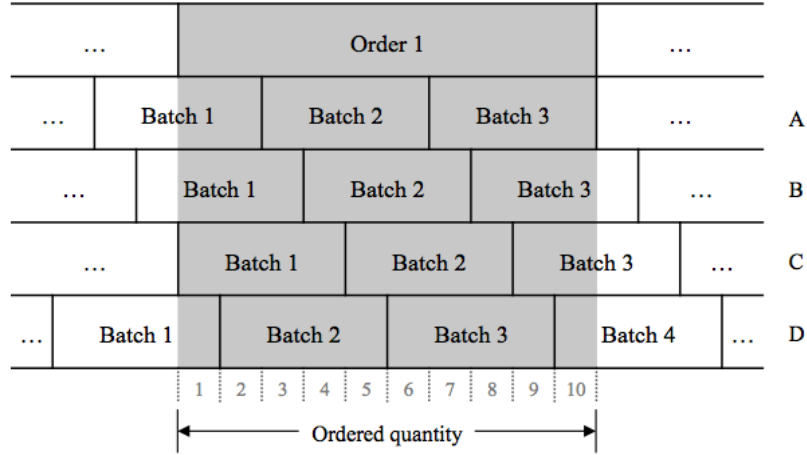

Figure 2. Possible fragmentations of an order of 10 units composed by batches of 4 units

Following the FIFO assignment policy, there are only four possible results, which are presented in Figure 2: in patterns A, B and C the order is composed by 3 batches; in pattern D the order is composed by 4 batches. Accordingly, the respective probabilities are ¾ = 0.75% chances of having 3 different batches in the order and ¼ = 0.25% chances of having 4.

The analysis of these possible combinations of order and batch sizes provides the following equations for calculating the probabilities of having a maximum and minimum number of fragmentations $P_{FrMAX}$ and $P_{FrMIN}$. In this case the multiplicity relationship between the order size and the batch size is a defining factor:

$$P_{FrMAX} \begin{cases} \text{If} \quad \mod\left(\dfrac{O_S}{B_S}\right) = 0 \quad \text{Then} \quad P_{FrMAX} = \dfrac{B_S - 1}{B_S} \\ \\ \text{If} \quad \mod\left(\dfrac{O_S}{B_S}\right) \neq 0 \quad \text{Then} \quad P_{FrMAX} = \dfrac{\mod\left(\dfrac{O_S}{B_S}\right) - 1}{B_S} \end{cases}$$

$$P_{FrMIN} = 1 - P_{FrMAX}$$

The number of fragments composing an order is then weighted as follows:

$$Fr_{BO} = P_{FrMIN} * Fr_{MIN} + P_{FrMAX} * Fr_{MAX}$$

In the given example, the number of fragments of batches (of 4 units) present in an order (of 10 units) is expected to be:

$$Fr_{4,10} = 0.75 * 3 + 0.25 * 4 = 3.25$$

Using the probability and boundary equations, the following expression is obtained in order to calculate the number of fragments of input batches present in an output order:

$$Fr_{BO}(O_S, B_S) \begin{cases} \text{If} \quad \mod\left(\dfrac{O_S}{B_S}\right) = 0 \quad \text{Then} \quad Fr_{BO} = \left\lfloor\dfrac{O_S}{B_S}\right\rfloor + \dfrac{B_S - 1}{B_S} \\ \\ \text{If} \quad \mod\left(\dfrac{O_S}{B_S}\right) \neq 0 \quad \text{Then} \quad Fr_{BO} = \left\lfloor\dfrac{O_S}{B_S}\right\rfloor + \dfrac{\mod\left(\dfrac{O_S}{B_S}\right) - 1}{B_S} \end{cases}$$

The general form of the fragments function is finally:

$$Fr_{BO}(O_S, B_S) = \frac{O_S + B_S - 1}{B_S}$$

This function shows that for a constant order size, the number of fragments composing the order will decrease as the batch sizes increase. Figure 3 shows this behaviour for a constant order size of 10 units and batch sizes varying from 1 to 20. The maximum number of fragmentations appears for a batch size equal to one and the minimum number of fragmentations leans towards an asymptote at $Fr_{BO} = 1$ for extremely large batch sizes.

Several batch sizing studies have shown that in the absence of setup costs, the optimal batch size is one (Zipkin, 1987; Karmarkar, 1987; Kekre, 1987; Lee 2006). It is important to note that **in terms of fragmentation the optimal choice is quite the opposite**. For a given order size, smaller input batches increase the likelihood of fulfilling the order with several different fragments. This will increase the probability of finding at least one unacceptable batch within the order, and consequently its inherent risk.

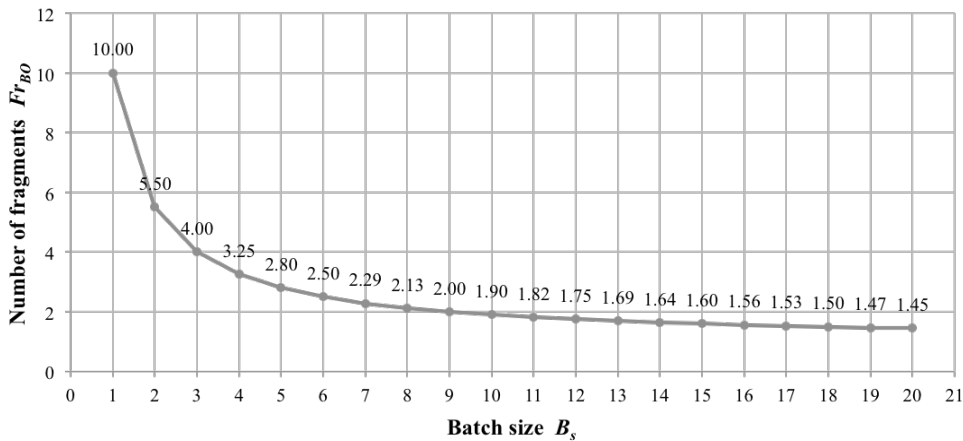

Figure 3. Number of fragments $Fr_{BO}$ present in a customer order of 10 units as a function of the batch size

In the previous example (cf. Figure 2), the function provides a result of 3.25 different batch fragments in each order ($B_S = 4$ in Figure 3). If any of these were to have a problem, the order would be recalled.

## 4.2. Probability of recalling an order

The chance of generating an unacceptable customer order is equal to the probability of having at least one unacceptable batch within the materials composing the order; that is, every case except for "all batches being acceptable". As the number of different batches composing the order is given by $Fr_{BO}$, such probability can be calculated as follows:

$$Pu_O = 1 - (1 - Pu_B)^{Fr_{BO}}$$

With:

$Pu_O$ = Probability of an unaccetable order

$Pu_B$ = Probability of an unaccetable batch = Probability of batch crisis

If the customer order concerns a production order, a mix of input material would be spoiled by the presence of an unacceptable batch, and the entire order would consequently be withdrawn.

## 4.3. Calculating the recall size

According to the equations presented in the previous sections, for a total fulfilled quantity the expected recall size can be calculated if the probability of having an unacceptable batch of a given input material is known.

The expected recall size "$R_S$" for a set of customer orders composing a total ordered quantity "$Q$" can be calculated in terms of the fragments of batches present in each order as follows:

$$R_S = f(Q, Pu_B, Fr_{BO}) = Q * \left(1 - (1 - Pu_B)^{Fr_{BO}}\right)$$

Thus the general expression for the expected recall size can be expressed in terms of the batch size and the order size as follows:

$$R_S = Q * \left(1 - (1 - Pu_B)^{\left(\frac{O_S + B_S - 1}{B_S}\right)}\right)$$

## 5. Numerical simulation

A Monte Carlo simulation was developed in order to validate the mathematical approach. The objective of the simulation was to measure the recalled quantities after fulfilling a set of orders with a given set of batches that might be affected by a known probability of crisis (the probability of finding an unacceptable batch $Pu_B$). The parameters of the simulation are: the batch size, the order size, the total ordered quantity, the probability of batch crisis and the current level of the first batch (as shown in Figure 2, it might have been partially consumed by previous orders).

Figure 4 presents the main steps of the simulation: 1) the orders generation, in which a set of orders is generated according to the total ordered quantity and the average order size; 2) the batches generation, in which, based on the batch size, a set of batches is created composing the total on-hand inventory, each batch having its own "crisis status" depending on the probability of batch crisis; 3) the FIFO assignment, after randomly defining the current consumption level of the first batch, each order is fulfilled with the on-hand inventory following a FIFO policy; 4) the recall measure, in which the crisis batches are identified and all orders containing such batches are counted as withdrawn; 5) the results averaging, in which after a significant set of iterations, the recall results are averaged so the total recalled quantity can be estimated for each "ordered size - batch size" couple.

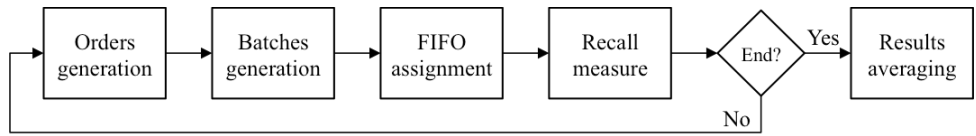

**Figure 4. Main steps of the simulation**

The crisis probability of the input batches is modelled following a uniform probability distribution. The current level of the first batch is randomly assigned, also following a uniform probability distribution.

Figure 5 presents the results of the simulation for an example of a total ordered quantity of 50 units that may be ordered in average order sizes varying from 1 to 50 units and fulfilled by on-hand batches of sizes varying from 1 to 100 units, each batch having a crisis probability of 15%. The value of 15% is chosen according to the work of Memon et al. (2014) for a low traceability level scenario. In order to guarantee a confidence interval of more than 98%, 10,000 iterations were performed for each set of parameters.

Some values of the recall landscape are easily identified, for example the lower right point of Figure 5 ($O_S = 1$; $B_S = 1$) indicates an average recall of 7.5 units when the total ordered quantity is made from 50 orders of size 1 and such orders are fulfilled by batches of size 1. In total there will be 50 batches each with a crisis probability of 15%. Each recalled order would affect only one unit. The recalled quantity is therefore easily calculated as 50 * 0.15 = 7.5.

Another interesting point appears on top of the figure ($O_S = 50$; $B_S = 1$), which indicates an average recall of 50 units (i.e. the total ordered quantity is recalled) when a single order of size 50 is fulfilled with size 1 batches. As a result, the order contains 50 different batches, and the probability of having at least one unacceptable batch is extremely close to 1.

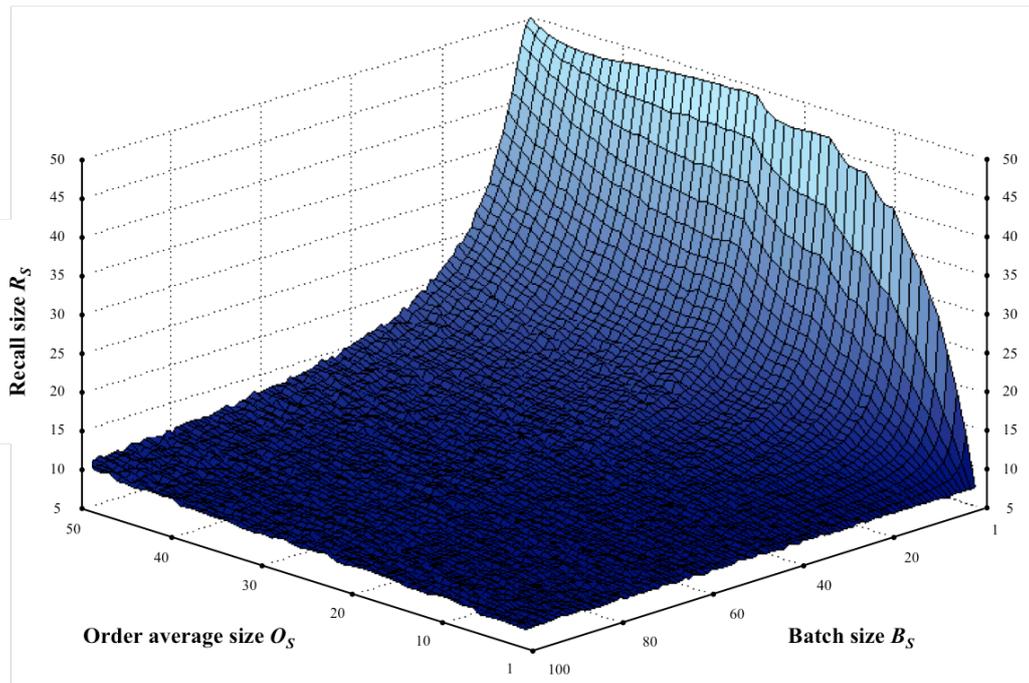

**Figure 5. Average recall results of the Monte Carlo simulation after 10,000 iterations for a probability of batch crisis $Pu_B$ = 15%**

These simulation results are compared against the recall sizes $R_S$ calculated with the mathematical model using the equation $R_S = f(Q, Pu_B, Fr_{BO})$. The average absolute error obtained is 1.463%, which validates the reliability of the mathematical approach.

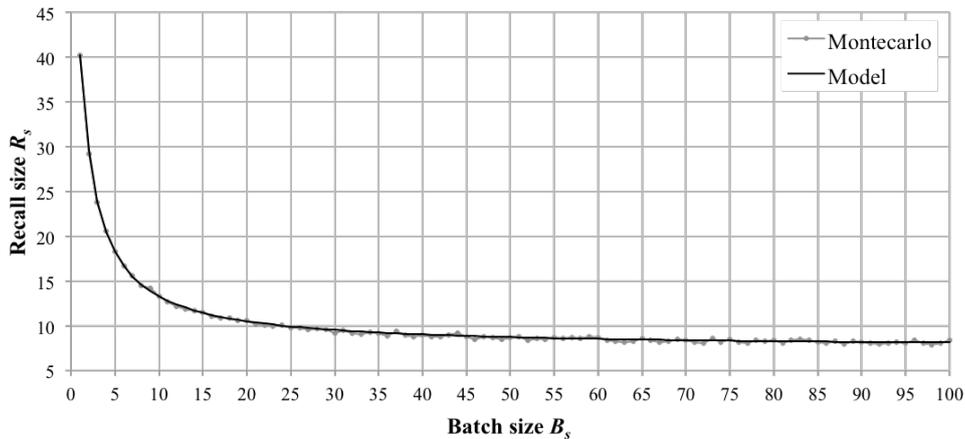

**Figure 6. Recalled quantity for a total ordered quantity $Q$ = 50; an order size $O_S$ = 10; and a probability of crisis $Pu_B$ = 15%; comparison between the results of the Monte Carlo simulation and the mathematical model**

Figure 6 illustrates the results for a fixed order size equal to 10. The grey dotted line represents the average size of the recall obtained with the Monte Carlo simulation, and the black plane line represents the theoretical values calculated with the mathematical model.

Other simulations were performed with different values of crisis probabilities $Pu_B$ and total ordered quantities $Q$. The order sizes varied from 1 to the total ordered quantity, and the batch sizes varied from 1 to double the total ordered quantity. The results validate the proposed mathematical model, as in all cases the resulting average error was less than 1.5%.

# 6. Discussion: best practices, contradictions and assertions

The proposed analysis allows a direct quantification of the impact that batch sizes of input materials and customer order sizes have on the product recall sizes. According to the internal and external customer-supplier relationships, some organizations might be able to adapt their batch sizing policies; others might be able to modify their customer's (internal or external) ordering policies; and some might even be able to do both.

## 6.1. Variable batch sizes and fixed customer orders

If the average customer order sizes are accepted as fixed but the batch sizes can be adapted, it is noticeable that larger input batch sizes will reduce the number of fragments present in the customer orders. For very big batch sizes (when the batch size tends towards infinite), the number of fragments present in the orders leans towards 1 and the recall size tends towards an asymptote in $R_S = Q * Pu_B$.

$$\lim_{B_S \to \infty}(R_S) = Q * \left(1 - (1 - Pu_B)^{\left(\frac{O_S + \infty - 1}{\infty}\right)}\right) = Q * Pu_B$$

Figure 7 illustrates this behaviour and makes it clear that to reduce the recalled quantities it is recommended to augment the batch sizes of input materials.

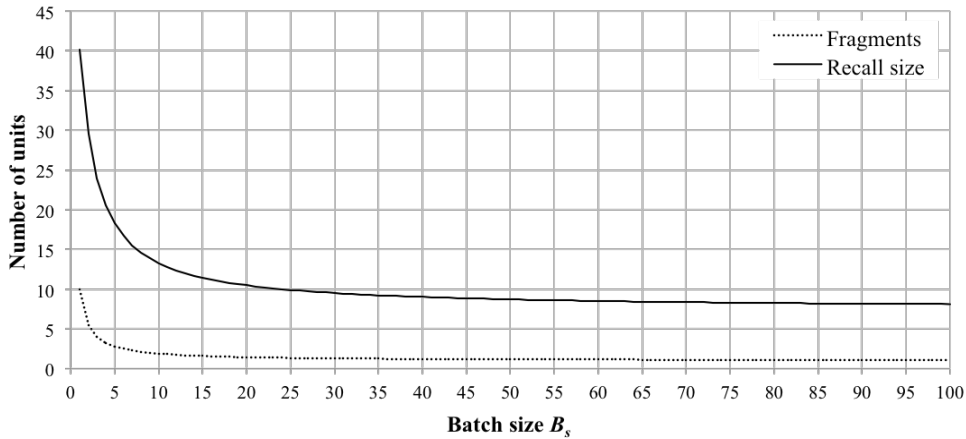

**Figure 7. Variation of the fragments $Fr_{BO}$ and recall size $R_S$ as functions of the input batch size $B_S$ for a fixed order size $O_S$ = 10, a probability of crisis $Pu_B$ = 15% and a total ordered quantity $Q$ = 50**

Smaller batches increase the mix on the customer orders and thus the spreading of the risk in case of crisis. A contradiction emerges between this statement and the flexibility policies that seek to produce and deliver using small batches (Diaby, 2000; Johnson and Business, 2003; Poon et al., 2011).

It is important to clarify that an input batch can be significantly large and yet be supplied in small ordering quantities. Hence this statement does not necessarily contradict the principles of pull-flow systems and inventory reduction, even though it does argue for the need for a controlled supply policy in which materials can be homogenized over time.

## 6.2. Fixed batch sizes and variable customer orders

If batch sizes are accepted as fixed but the customer order sizes can be influenced (for instance, in the production setting the customer orders represent the production lots, which can be modified), it is evident that large order sizes will increase the chances of having a composition made of several batches. For very big order sizes (when the order size tends towards infinite), the number of fragments present in the orders leans towards infinite and the recall size tends towards an asymptote in $R_S = Q$, i.e. all produced or shipped quantities are recalled.

$$\lim_{O_S \to \infty}(R_S) = Q * \left(1 - (1 - Pu_B)^{\left(\frac{\infty + B_S - 1}{B_S}\right)}\right) = Q$$

By contrast, smaller order sizes will reduce the number of batch fragments, as shown in Figure 8.

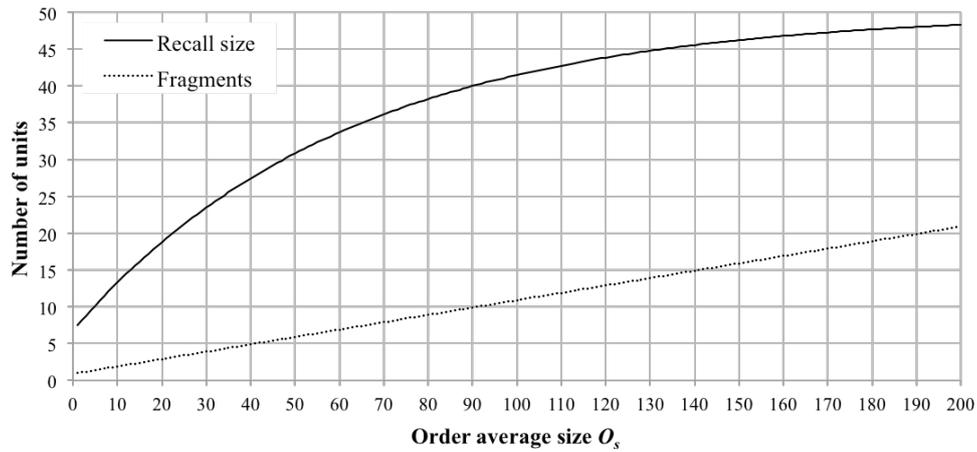

**Figure 8. Variation of the fragments $Fr_{B0}$ and recall size $R_S$ as functions of the order size $O_S$ for a fixed input batch size $B_S$ = 10, a probability of crisis $Pu_B$ = 15% and a total ordered quantity $Q$ = 50**

It is clear that from a customer-oriented standpoint, the aim of organizations in terms of risk is to work alongside clients in order to minimize the order sizes and consequently the dispersion of risk. This objective of reducing customer order sizes is consistent with the actual trends of pull-flow and highly frequent ordering systems.

## 6.3. Overall findings

As discussed in previous sections, **from a risk-oriented perspective the overall objective is to transform using large batches of input material in order to fulfil small customer orders**. Nonetheless, some recommendations have a larger impact on the recall size function. Figure 9 presents the variations in the shape of the recall size function for different batch crisis probabilities.

It is important to note that unitary orders ($O_S$ = 1) always result in a minimum recall size $R_S = Q*Pu_B$ and are independent of the batch size variation. By contrast, unitary batches ($B_S$ = 1) mainly result in large recalls that vary depending on the order sizes, for larger orders increase the recall size. A unitary batch will engender a small recall only if the order size is also unitary; otherwise the order size will be greater than the batch size and will unavoidably be fulfilled with several fragments.

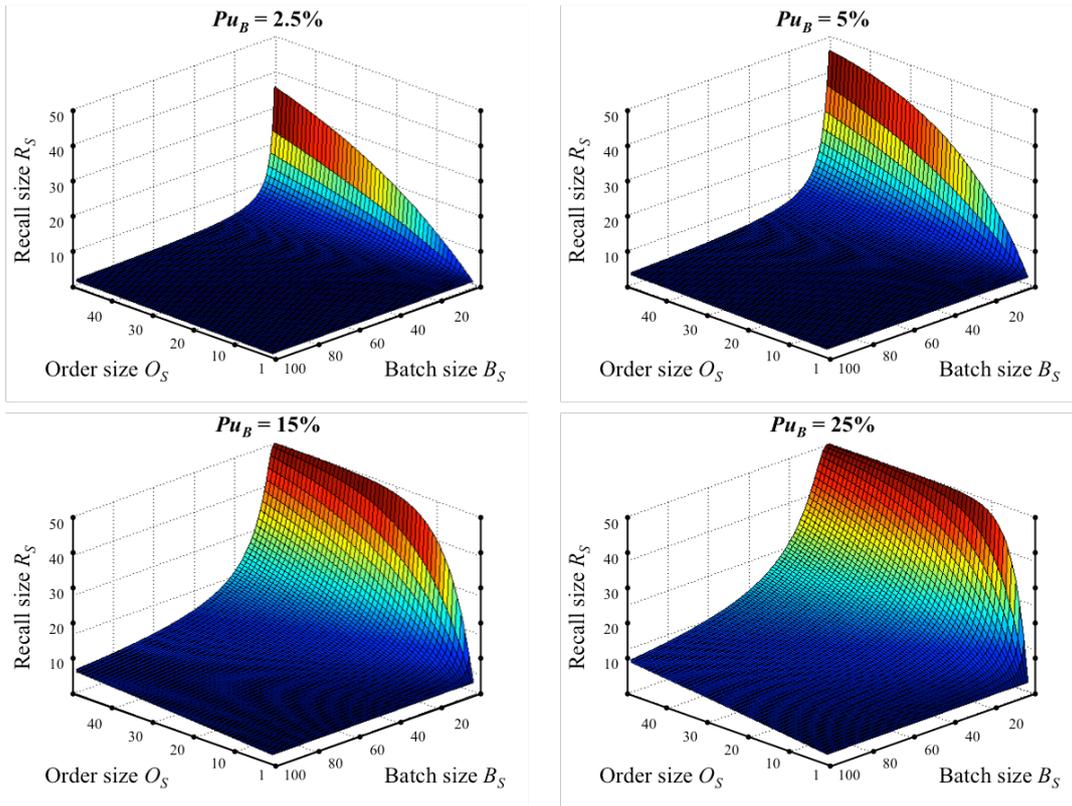

**Figure 9. Variation of the recall size function $R_S$ shape for different batch crisis probabilities $Pu_B$ and total ordered quantity of $Q = 50$**

Larger batch sizes generally result in small recalls. These are also increased by larger orders, but such impact is marginal for crisis probabilities under 25% (cf. Figure 9) and will only be significant for higher crisis probabilities.

Table 1 summarizes some significant relations of the recall size function:

|  | $B_S = 1$ | $B_S \leq x$ | $B_S > x$ | $B_S \approx \infty$ |
|---|---|---|---|---|
| $O_S = 1$ | Minimum $R_S$ $R_S = Q*Pu_B$ | | | |
| $O_S = x$ | Very high $R_S$ | High $R_S$ | Moderate $R_S$ | Minimum $R_S$ $R_S = Q*Pu_B$ |
| $O_S \approx \infty$ | Highest $R_S$ $R_S = Q$ | Highest $R_S$ $R_S = Q$ | Highest $R_S$ $R_S = Q$ | Low $R_S$ $R_S = Q*(1-(1-Pu_B)^2)$ |

**Table 1. Recall size results for significant values of the batch size and order size**

Smaller input batch sizes are the foremost obstacle to overcome while minimizing the recall size function. Big orders composed by small batches (highest fragmentation) are entirely recalled, but even middle-sized orders, in relation to the total ordered quantity, result in large recalls when fulfilled with small batches. A noteworthy change in the behaviour of the recall function appears when the batch size is equal to or greater than the order size (cf. Figure 3 and Figure 7), since starting from this point the order can be fulfilled with one single batch. This is seldom the case when the two quantities are very close, but it becomes a broader rule when the batch size is significantly larger than the order size. In idealistic conditions the batch sizes should be as large as possible but in the lesser acceptable cases they should be at least greater than the average order sizes.

# 7. Conclusions

This analysis investigates the interactions amongst recall sizes, batch sizes, order sizes and fragmentations of input materials, within a FIFO assignment policy. Batch splitting and dispersion are critical factors in traceability procedures and it has been proved that they affect the results of product recall procedures. In this paper, for the first time, the notion of batch fragments is formally related to the recall size as an amplifying factor.

A mathematical approach is presented in order to calculate the expected number of fragments of different input batches present in a customer order. The number of fragments depends on both the sizes of the on-hand materials batches and the average sizes of the customer orders. As an example, in the production setting the proposed approach allows the calculation of the expected number of fragments of batches of a raw material present in the average production of a finished product.

If at least one of the batches composing a customer order is unacceptable (i.e. it does not satisfy the quality/safety standards), the order should be withdrawn. The probability of having an unacceptable batch, (namely, the batch crisis probability) can be known and therefore allows the calculation of the recall size for a given ordered quantity. We therefore propose an equation in order to calculate the recall size resulting from a known crisis probability of an input material used (and fragmented) to fulfil a set of customer orders that satisfy a total ordered quantity.

An exponential relationship has been found between the number of fragments and the recall size in case of batch crisis. The mathematical model has been validated using a Monte Carlo simulation.

The results indicate that **the ideal conditions minimizing the recall size appear for large batches of input materials and small sizes of customer orders**. In other words, organizations should use large batches of inputs to fulfil small orders of outputs. Clearly, in terms of risk it is preferable to work with larger input batches as this reduces the probability finding multiple fragments in the customer orders, although larger batches engender higher stock levels and reduce the overall flexibility of operations. It appears that a compromise needs to be reached in terms of upstream batch sizing policies. The idea of fulfilling small customer orders is nonetheless coherent with the trends on pull-flow systems, which aim to produce and deliver in small quantities. It is in any case advised to avoid configurations in which batch sizes are smaller than average order sizes, as these have been identified as zones of risk amplification.

These conclusions show the advantages for organizations to homogenize their inputs and narrow down their outputs in order to minimize the propagation of risks. Such strategy stimulates a volume and cost-oriented management of the upstream flow of input materials and a just-in-time pull-oriented management of the downstream customer orders.

Several questions remain open for future research:

- This analysis has been proposed for a single input material that is transformed or used to fulfil customer orders, but such orders can be composed of several input materials. Although the proposed approach is valid for studying one critical component that has a known crisis probability (for instance a meat component in a food production setting), this type of analysis can be extended in order to consider the simultaneous effect of different input materials, each with its own batch configuration and intrinsic risk. Moreover, the relation between the batch crisis probability and the recall size remains to be studied.

- In this work, the probability of crisis of a batch is considered to be independent from the batch size. This is a basic approach that derives from the definition of batch (ECDPH, 2004). However, this assumption could be questioned and specified by studying the empirical relations between the probabilities of crisis of a set of batches and their sizes.

- The proposed model defines the number of fragments of input batches for average values of order sizes. This is valid for the production setting in which input batches of raw materials are used in order to make uniform quantities of finished products (the production lots). However, in the distribution setting the sizes of the customer orders vary, following a probability distribution (typically Gaussian). If such variation is important (i.e. high standard deviations) it should have a significant impact on the fragmentation result, which remains to be modelled.